\documentclass[11pt,letterpaper]{article}

\usepackage{latexsym,multirow}
\usepackage{amssymb,amsmath, bm}
\usepackage{graphicx}

\usepackage[color,all,import,arrow]{xy}

\usepackage{enumerate}

\usepackage{natbib}
\usepackage[pdftex, bookmarksopen=true, bookmarksnumbered=true,
pdfstartview=FitH, breaklinks=true, urlbordercolor={0 1 0}, citebordercolor={0 0 1}]{hyperref}
\usepackage{colortbl}
\usepackage{subfigure}

\usepackage{tkz-graph}  
\usetikzlibrary{shapes.geometric}%

\usepackage{dcolumn}
\newcolumntype{.}{D{.}{.}{-1}}
\newcolumntype{d}[1]{D{.}{.}{#1}}
\usepackage{theorem}
\theoremstyle{plain}
\theoremheaderfont{\scshape}

\newcommand{\ind}{\mbox{$\perp\!\!\!\perp$}}

\usepackage{rotating}

\usepackage[compact]{titlesec}

\allowdisplaybreaks

\newcommand\spacingset[1]{\renewcommand{\baselinestretch}%
{#1}\small\normalsize}

\newcommand{\blind}{0}

\newcommand{\btheta}{\boldsymbol{\theta}}

\newcommand{\bZ}{\bm{Z}}

\newcommand{\bV}{\mathbf{V}}
\newcommand{\bz}{\bm{z}}

\newcommand{\bA}{\bm{A}}
\newcommand{\cA}{\mathcal{A}}
\newcommand{\ba}{\bm{a}}

\newcommand{\E}{\mathbb{E}}
\newcommand{\bX}{\mathbf{X}}
\newcommand{\bx}{\mathbf{x}}

\newcommand{\bW}{\mathbf{W}}

\begin{document} 

\newcommand{\tit}{Discussion of ``The Blessings of Multiple Causes'' by Wang and Blei}

%
%
\spacingset{1.25}

\if0\blind

{\title{\bf\tit\thanks{We thank Naoki Egami, Connor Jerzak, Michael
      Li, and Xu Shi for helpful discussions.}}

  \author{Kosuke Imai\thanks{Professor, Department of Government and
      Department of Statistics, Institute for Quantitative Social
      Science, Harvard University, Cambridge MA 02138. Phone:
      617--384--6778, Email:
      \href{mailto:Imai@Harvard.Edu}{Imai@Harvard.Edu}, URL:
      \href{https://imai.fas.harvard.edu}{https://imai.fas.harvard.edu}}
    \hspace{.5in} Zhichao Jiang\thanks{Assistant Professor, Department
      of Biostatistics and Epidemiology,
      University of Massachusetts, Amherst MA 01002.  }  }

\date{
\today
}

\maketitle

}\fi

\if1\blind
\title{\bf \tit}

\maketitle
\fi

\pdfbookmark[1]{Title Page}{Title Page}

\thispagestyle{empty}
\setcounter{page}{0}

%
%
%


\spacingset{1.5}

We begin by congratulating Yixin Wang and David Blei for their
thought-provoking article that opens up a new research frontier in the
field of causal inference.  The authors directly tackle the
challenging question of how to infer causal effects of many treatments
in the presence of unmeasured confounding.  We expect their article to
have a major impact by further advancing our understanding of this
important methodological problem.  This commentary has two goals.  We
first critically review the deconfounder method and point out its
advantages and limitations.  We then briefly consider three possible
ways to address some of the limitations of the deconfounder method.

\section{The Advantages and Limitations of the Deconfounder Method}

We first discuss several advantages offered by the deconfounder
method.  We then examine the assumptions required by the method and
discuss its limitations.

\subsection{The Deconfounder Method}

Suppose that we have a simple random sample of $n$ units from a
population.  We have a total of $m$ treatments, represented by the
$m$-dimensional vector, $\bA_i=(A_{i1}, A_{i2}, \ldots, A_{im})^\top$,
for unit $i$.  For the sake of simplicity, we ignore the possible
existence of observed confounders $\bX_i$.  But, all the arguments of
this commentary are applicable, conditional on $\bX_i$.  The
deconfounder method consists of the following simple two steps.  The
first step fits the following factor model to the observed treatments,
\begin{equation}
  p(A_{i1}, A_{i2}, \ldots, A_{im}) \ = \  \int p(\bZ_i) \prod_{j=1}^m p(A_{ij}
  \mid \bZ_i) \ d\bZ_i, \label{eq:factor}
\end{equation}
where $\bZ_i=(Z_{i1}, Z_{i2}, \ldots, Z_{ik})^\top$ represents the
$k$-dimensional vector of latent factors.

Once the estimates of the factors $\widehat{\bZ}_i$, which Wang and
Blei call the {\it substitute confounders}, are obtained, the second
step estimates the average causal effects of multiple treatments by
adjusting for these substitute confounders as follows,
\begin{equation}
\tau(\ba, \ba^\prime) \ = \ \E\{Y_i(\ba) - Y_i(\ba^\prime)\} \ = \ \E\{\E(Y_i \mid \bA_i = \ba,
\widehat{\bZ}_i) - \E(Y_i \mid \bA_i = \ba^\prime, \widehat{\bZ}_i)\}, \label{eq:reg}
\end{equation}
where $\ba \in \cA$ and $\ba^\prime \in \cA$ are the vectors of
selected treatment values with $\ba \ne \ba^\prime$ and $\cA$
represents the support of $\bA_i$.  In practice, a regression model
may be used to adjust for the substitute confounders as demonstrated
by Wang and Blei in their empirical application.

The deconfounder method is attractive to applied researchers for
several reasons.  First, it is a simple procedure based on two classes
of familiar statistical models --- factor models and regression
models.  Second, the method offers diagnostics in observational
studies with unmeasured confounding.  Specifically, researchers can
check the conditional independence among the observed treatments given
the estimated factors,
\begin{equation}
  A_{ij} \ \ind \ \bA_{i,-j} \mid \widehat{\bZ}_i
\end{equation}
for any $j=1,\ldots,m$ and $\bA_{i,-j}$ represents all the treatments
except $A_{ij}$.  If this conditional independence does not hold, then
there may exist unobserved confounders that affect both $A_{ij}$ and
some of $\bA_{i,-j}$, yielding a biased causal estimate.  As discussed
below, however, the lack of conditional independence may also be due
to the misspecification of factor model, which, for example, would be
present if there are causal relationships among treatments.

In sum, the deconfounder method proposes a simple solution to a
long-standing problem of inferring causal effects of multiple
treatments in observational studies.  Many analysts of observational
studies rely upon the assumption that the treatments are unconfounded
conditional on a set of observed pre-treatment covariates.  And yet,
it is often difficult to rule out the possible existence of unobserved
confounders.  The deconfounder method not only offers a new
identification strategy in the presence of unobserved confounding, but
also shows how to check the validity of the resulting estimates under
certain assumptions.

\subsection{Assumptions}

\begin{figure}[t]
  \begin{center}
    \tikzstyle{VertexStyle} = [shape = circle, minimum width = 2ex, draw]
    \tikzstyle{EdgeStyle}   = [->,>=stealth']      
    \begin{tikzpicture}[scale=1] 
      \SetGraphUnit{2}
      \node[VertexStyle] (A1) at (0, 0) {$A_1$}; 
      \node[VertexStyle] (A2) at (2, 0) {$A_2$}; 
      \node[VertexStyle] (Am) at (6, 0) {$A_m$}; 
      \node[VertexStyle] (Y) at (3, 2) {$Y$}; 
      \node (dots) at (4, 0) {$\cdots$};
      \node[circle, dashed, draw] (Z) at (3, -2) {$\bZ$};
      \Edges(A1, Y) \Edges(A2, Y) \Edges(Am, Y)
      \Edges(Z, A1) \Edges(Z, A2) \Edges(Z, Am) \Edges(Z, Y)      
    \end{tikzpicture}
  \end{center}
  \vspace{-.25in}
  \caption{Directed Acyclic Graph for the Deconfounder Method.} \label{fig:DAG}
\end{figure}
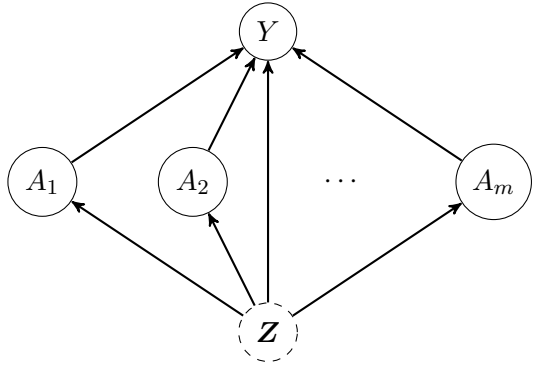

What assumptions does the deconfounder method require?  Wang and Blei
uses a graphical model to represent the conditional dependencies
required by the deconfounder method.  Here, we reproduce the graphical
model using the directed acyclic graph (DAG) in Figure~\ref{fig:DAG}.
In addition to the SUTVA \citep{rubi:90}, this DAG implies several key
assumptions.  First, the unobserved confounders $\bZ$ should represent
all confounding variables such that the treatments are ignorable given
$\bZ$,
\begin{equation}
  Y_i(\ba) \ \ind \ \bA_i \mid \bZ_i \label{eq:unconfounded}
\end{equation}
for any $\ba \in \cA$.  The assumption implies that the multi-cause
confounder $\bZ_i$ suffices to adjust for the treatment-outcome
confounding.

Second, the DAG also implies the following conditional independence
assumption,
\begin{equation}
  A_{ij} \ \ind \ \bA_{i,-j} \mid \bZ_i \label{eq:mutual_cause}
\end{equation}
for any $j=1,2,\ldots,m$. The assumption justifies the factor model in
equation~\eqref{eq:factor}.  This assumption is violated if, for
example, there exists a causal relationship among treatments.  In the
movie revenue application considered in the original article, the
assumption is violated if the choice of actor for the main role (e.g.,
Sean Connery in a James Bond movie) influences the selection of actor
for another role (e.g., Bernard Lee as the character of M).  This is
an important limitation of the deconfounder method as the problem may
be common in applied research with multiple treatments.

In addition, according to Wang and Blei, the deconfounder method also
requires the following overlap assumption that is not explicitly
represented in the DAG,
\begin{equation}
  p(\bA_i \in \cA^\ast \mid \bZ_i) \ > \ 0 \label{eq:overlap}
\end{equation}
for all sets $\cA^\ast \subset \cA$ with $p(\bA_i \in \cA^\ast) > 0$.
The assumption implies that the choice of treatment values $\ba$ may
be constrained when estimating $\E\{Y_i(\ba)\}$.  If the selected
value of $\ba$ does not belong to $\cA^\ast$, then the resulting
causal inference will be based on extrapolation.

Finally, the key identification condition of the deconfounder method
is the assumption of ``no unobserved single-cause confounder.''  Wang
and Blei formalize this assumption as the following set of conditional
independence assumptions (see Definition~4 of the original article),
\begin{eqnarray}
  Y_i(\ba) & \ind & A_{ij} \mid \bV_{ij} \label{eq:single} \\
  A_{ij} & \ind & \bA_{i,-j} \mid \bV_{ij} \label{eq:indep}
\end{eqnarray}
for any $j=1,2,\ldots,m$, $\ba \in \cA$, and some random variable
$\bV_{ij}$.  In addition, the authors require that these conditional
independence relations do not hold when conditioning on any proper
subset of the sigma algebra of $\bV_{ij}$.

\begin{figure}[t]
  \vspace{-.25in}
  \spacingset{1}
  \begin{center}
  \subfigure[only unobserved single-cause confounders exist]{
    \tikzstyle{VertexStyle} = [shape = circle, minimum width = 2ex, draw]
    \tikzstyle{EdgeStyle}   = [->,>=stealth']      
    \begin{tikzpicture}[scale=1] 
      \SetGraphUnit{2}
      \node[VertexStyle] (A1) at (1, 0) {$A_1$}; 
      \node[VertexStyle] (A2) at (3, 0) {$A_2$}; 
      \node[VertexStyle] (A3) at (5, 0) {$A_3$}; 
      \node[VertexStyle] (Y) at (3, 2) {$Y$}; 
      \node[circle, dashed, draw] (Z1) at (1, -2) {$\bZ_1$}; 
      \node[circle, dashed, draw] (Z2) at (3, -2) {$\bZ_2$}; 
      \node[circle, dashed, draw] (Z3) at (5, -2) {$\bZ_3$}; 
      
     \draw [->, >=stealth', thick=2] (Z2) to [out=45, in=-75] (Y); 

      \Edges(A1, Y) \Edges(A2, Y) \Edges(A3, Y) 
      \Edges(Z1, A1) \Edges(Z2, A2) \Edges(Z3, A3) \Edges(Z1, Y)
      \Edges(Z3, Y)
    \end{tikzpicture}
  }
  \hspace{.5in}
\subfigure[both unobserved single-cause and multiple-cause confounders exist]{
    \tikzstyle{VertexStyle} = [shape = circle, minimum width = 2ex, draw]
    \tikzstyle{EdgeStyle}   = [->,>=stealth']      
    \begin{tikzpicture}[scale=1] 
      \SetGraphUnit{2}
      \node[VertexStyle] (A1) at (1, 0) {$A_1$}; 
      \node[VertexStyle] (A2) at (3, 0) {$A_2$}; 
      \node[VertexStyle] (A3) at (5, 0) {$A_3$}; 
      \node[VertexStyle] (Y) at (3, 2) {$Y$}; 
      \node[circle, dashed, draw] (Z1) at (1, -2) {$\bZ_1$}; 
      \node[circle, dashed, draw] (Z2) at (3, -2) {$\bZ_2$}; 
      \node[circle, dashed, draw] (Z3) at (5, -2) {$\bZ_3$}; 
      
     \draw [->, >=stealth', thick=2] (Z2) to [out=60, in=-75] (Y); 

      \Edges(A1, Y) \Edges(A2, Y) \Edges(A3, Y) \Edges(Z2, A1)
      \Edges(Z2, A3)
      \Edges(Z1, A1) \Edges(Z2, A2) \Edges(Z3, A3) \Edges(Z1, Y)
      \Edges(Z3, Y)
    \end{tikzpicture}
}
\end{center}
\vspace{-.2in}
  \caption{Examples of Unobserved Single-cause Confounders.} \label{fig:DAG2}
\end{figure}
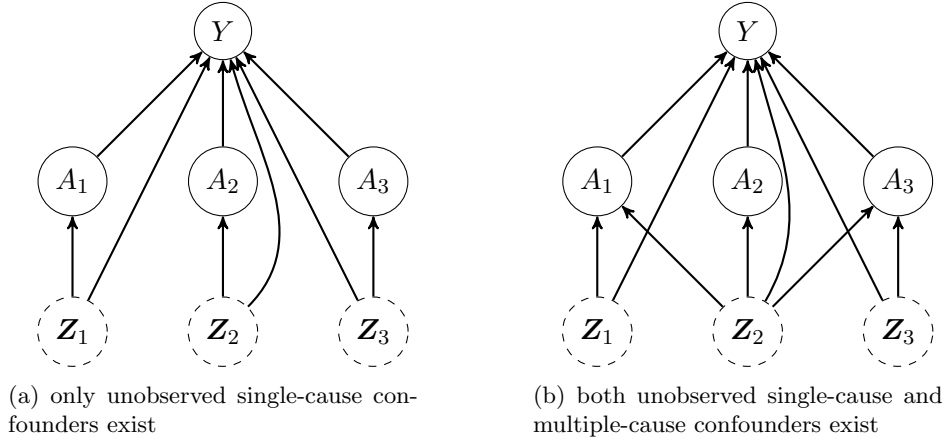

Unfortunately, these conditional independence assumptions are not
sufficient to eliminate the possible existence of unobserved
single-cause confounders.  Figure~\ref{fig:DAG2} presents two
examples, in which single-cause confounders exist, but
equations~\eqref{eq:single}~and~\eqref{eq:indep} still hold.  In
addition, both cases can be reduced to the DAG in Figure~\ref{fig:DAG}
where no single-cause unobserved confounder exists by defining the
unobserved multi-cause confounder as $\bZ = (\bZ_1, \bZ_2, \bZ_3)$.
The examples demonstrate that a single multi-cause confounder can be
decomposed into multiple single-cause confounders, and that several
single-cause confounders can be combined into a single multi-cause
confounder. Therefore, it is difficult to distinguish between
single-cause and multiple-cause confounders without the knowledge of
causal relationships among the variables.

We believe that it is important to develop the precise formal
statement of the no unobserved single-cause confounder assumption.
Such formalization allows us to understand how this assumption enables
the identification of causal effects.  In addition, our discussion
implies that assessing the credibility of the assumption requires the
scientific knowledge about the underlying causal structure involving
unobserved confounders.

\subsection{Nonparametric Identification}

Wang and Blei establish the nonparametric identification of the
average treatment effect given in equation~\eqref{eq:reg} under the
aforementioned assumptions in two steps.  First, they show that a
factor model of the observed treatments can be used to consistently
estimate the substitute confounder.  Second, they show that given the
substitute confounder, the average treatment effects can be
nonparametrically identified using equation~\eqref{eq:reg} above.

In an insightful paper, \citet{d2019multi} demonstrates that this
two-step proof strategy leads to two problems for the deconfounder
method.  First, there may be more than one factor model that is
compatible with the distribution of the observed treatments.  He
provides an example where different factor models that are compatible
with the distribution of the observed treatments under the structure
of Figure~\ref{fig:DAG} yield different causal estimates.  Second,
\citeauthor{d2019multi} shows that even if a factor model is uniquely
identified, the nonparametric identification is in general
impossible.

Moving beyond the counterexamples, we consider the identification
assumption for the factor model, discuss the role of the substitute
confounder, and assess the overlap assumption required by the
deconfounder method.

With respect to the identifiability of factor models,
\citet{kruskal1977three} and \citet{allman2009identifiability} give
the general identification assumptions when observed variables are
discrete.  In this case, a crucial assumption is that the latent
factor is correlated with the observed variables.  In our context,
this means that $\bZ$ must causally affect each treatment $A_j$.  In
the causal inference literature, this assumption is known as
faithfulness \citep{spirtes2000causation}, which states that there
exists conditional independence among variables in the population
distribution if and only if it is entailed in the corresponding DAG.
Thus, although Wang and Blei only discuss a set of conditional
independence assumptions, the deconfounder method requires the
faithfulness assumption in order to ensure the identifiability of
factor model.

Next, we discuss the role of the substitute confounder.  In the proof
of the deconfounder method, Wang and Blei not only assume that the
true unobserved confounder $\bZ_i$ can be consistently estimated, but
also treat the estimated substitute confounder $\widehat{\bZ}_i$ as
its true counterpart.  This proof strategy ignores the crucial fact
that the (estimated) substitute confounder is a function of observed
treatments
$\widehat{\bZ}_i=\widehat{h}_M(\bA_i)=\E_M(\bZ_i\mid \bA_i)$, where
$\hat{h}_M$ indicates the fact that the substitute confounder is
estimated from the data and depends on the choice of factor model and
$\E_M$ represents the expectation with respect to the fitted factor
model.  We emphasize that the substitute confounder $\widehat{\bZ}_i$
does not converge in probability to the true confounder $\bZ_i$, which
in itself is a random variable.  Rather, the substitute confounder
converges to a function of observed treatments.  Yet, this consistency
result is required for the key results of the paper (i.e.,
Theorems~6--8).

We also closely examine the identification formula given in
equation~\eqref{eq:reg} by explicitly writing out the conditional
expectation,
\begin{eqnarray}
 \ \E\{\E(Y_i \mid \bA_i = \ba,
\widehat{\bZ}_i)\} =\int \E(Y_i \mid \bA_i = \ba, \widehat{\bZ}_i)
  p(\widehat{\bZ}_i)  d\widehat\bZ_i \label{eq:deconfounder}
\end{eqnarray}
Notice that equation~\eqref{eq:deconfounder} does not follow unless
the support of $p(\widehat{\bZ}_i \mid \bA_i = \ba)$ is identical to
the support of $p(\widehat{\bZ}_i)$ for any given $\ba \in \cA$.
Unfortunately, since the substitute confounder is estimated using the
observed treatments, $p(\widehat{\bZ}_i \mid \bA_i = \ba)$ is in
general degenerate.  The overlap assumption given in
equation~\eqref{eq:overlap} is not applicable because the assumption
is about the (true) unobserved confounders $\bZ_i$ rather than the
(estimated) substitute confounders, $\widehat{\bZ}_i$.  This means
that we can only identify
$\E(Y_i \mid \bA_i = \ba, \widehat{\bZ}_i = \bz)=\E(Y_i \mid \bA_i =
\ba)$ for the values of $\bz$ with $\bz = \widehat{h}_M(\ba)$,
implying that only a certain set of causal effects are identifiable. 

In Theorem~6 of the original paper, Wang and Blei address this problem
by imposing two additional restrictions.  First, it is assumed that
the outcome is separable in the following sense,
\begin{eqnarray}
  \E\{Y_i(\ba) \mid \widehat{\bZ}_i\} & = & f_1(\ba) + f_2(\widehat{\bZ}_i), \label{eq:separable1}\\
  \E(Y_i \mid \bA_i, \widehat{\bZ}_i) & = & f_3(\bA_i) + f_4(\widehat{\bZ}_i), \label{eq:separable2}
\end{eqnarray}
where we use $\widehat{\bZ}_i$ instead of $\bZ_i$ to emphasize the
fact that the substitute confounder is estimated.  Although
equation~\eqref{eq:separable1} allows us to write the average
treatment effect as a function of treatment values alone, i.e.,
$\E\{Y_i(\ba) - Y_i(\ba^\prime)\} = f_1(\ba) - f_1(\ba^\prime)$, this
assumption is not particularly helpful for identification since
conditioning on $\widehat{\bZ}_i$ is still required to identify the
mean potential outcomes.  In addition, equation~\eqref{eq:separable2}
can be rewritten as
$\E(Y_i \mid \bA_i) = f_3(\bA_i) + f_4(\hat{h}_M(\bA_i))$ because
$\widehat{\bZ}_i$ is a deterministic function of $\bA_i$.  This
suggests that the validity of this restriction about the outcome model
critically depends on the choice of factor model.

The second restriction is that when the treatments are continuous, the
substitute confounder is a piece-wise constant function, i.e.,
$\nabla_{\ba} f_{\btheta}(\ba)= 0$ where a parametric model is assumed
for
$p(\widehat{\bZ}_i \mid \bA_i = \ba, \btheta) =
\delta_{f_{\btheta}(\ba)}$ with a vector of parameters $\btheta$.  A
similar restriction is proposed for the case of discrete
treatments. Since
$p(\widehat{\bZ}_i \mid \bA_i = \ba, \btheta) =
\delta_{\hat{h}_M(\ba)}$ automatically holds, the assumption is valid
if $\hat{h}_M(\ba)$ is a piece-wise constant function.  Thus, this
second restriction also suggests that the choice of factor model is
critical for the validity of the deconfounder method.

In sum, we conclude that the nonparametric identification is generally
difficult to obtain under the deconfounder method. Because the
substitute confounder is a function of observed treatments, it leads
to the violation of the overlap assumption.  Wang and Blei introduce
two additional restrictions to address this problem.  However, these
assumptions impose severe constraints on the choice of factor model as
well as that of outcome model. As a consequence, they may
significantly limit the practical applicability of the deconfounder
method.  Even when researchers carefully choose a factor model that
satisfies these restrictions, they may obtain causal effects only for
a restricted range of treatment values.

\section{Alternative Approaches}

We next consider three alternative approaches to the important
question of identifying the causal effects of multiple treatments in
the presence of unobserved confounders.  The approaches in this
section will be based on equation~\eqref{eq:unconfounded}. Unlike the
deconfounder method, however, we will directly consider the
identification of the probability distributions involving the (true)
unobserved confounder $p(\bA_i, \bZ_i)$ and
$p(Y_i \mid \bA_i, \bZ_i)$ rather than adopting Wang and Blei's
two-step proof strategy.

\subsection{Parametric Approach}

Wang and Blei use parametric models in their empirical applications.
Here, we consider a more general parametric approach.  A primary
advantage of the parametric approach is simplicity, whereas its major
limitation is the required modeling assumptions that may not be
credible in practice.

Suppose that there exists a uniquely identifiable factor model for the
treatments, and that the joint distribution of $(\bA, \bZ)$ is also
identifiable.  We assume the following additive model for the outcome
variable,
\begin{eqnarray*}
  \E\{Y_i(\ba) \mid \bZ_i\} \ = \ \sum_{j=1}^m \beta_j b_j(a_j)+ \sigma g(\bZ_i),
\end{eqnarray*}
where $b_j(\cdot)$ and $g(\cdot)$ are pre-specified functions.  Under
this setting, it can be shown that if $\sigma$ is known, then the
average treatment effect is identifiable so long as
$(b_1(A_{i1}),\ldots,b_m(A_{im}))$ is linearly independent.  In
contrast, if $\sigma$ is unknown, then the average treatment effect is
identifiable if
$(b_1(A_{i1}),\ldots,b_m(A_{im}), \E\{g(\bZ_i)\mid \bA_i\})$ is
linearly independent.  This linear independence assumption is
analogous to the overlap assumption discussed earlier, but the
assumption can be tested using the observed data.

To illustrate this parametric approach, consider an example, in which
we have three binary treatments $m=3$ and one binary latent factor
$Z_i$.  Further assume that we have the following outcome model,
\begin{eqnarray*}
 \E\{Y_i(\ba) \mid Z_i\} \ = \ \beta_0+ \sum_{j=1}^3 \beta_j A_{ij}+ \sigma Z_i.
\end{eqnarray*}
Now, consider a scenario, under which $A_{ij}$'s are mutually
independent of one another given $Z_i$.  Then, the joint distribution
$p(A_{i1}, A_{i2}, A_{i3}, Z_i)= p(Z_i)\prod_{j=1}^3 p(A_{ij} \mid
Z_i)$ is identifiable based on the joint distribution of
$(A_{i1}, A_{i2}, A_{i3})$ up to label switching
\citep[see][]{kruskal1977three}.  Note that the average treatment
effects are invariant to label switching.  Thus, under this condition,
even if $\sigma$ is unknown, $\beta_j$'s are identifiable so long as
$\E(Z_i \mid A_{i1},A_{i2},A_{i3})$ is not linear in
$(A_{i1},A_{i2},A_{i3})$.

\begin{figure}[t]
  \begin{center}
    \tikzstyle{VertexStyle} = [shape = circle, minimum width = 2ex, draw]
    \tikzstyle{EdgeStyle}   = [->,>=stealth']      
    \begin{tikzpicture}[scale=1] 
      \SetGraphUnit{2}
      \node[VertexStyle] (A1) at (0, 0) {$A_1$}; 
      \node[VertexStyle] (A2) at (2, 0) {$A_2$}; 
      \node[VertexStyle] (A3) at (4, 0) {$A_3$}; 
      \node[VertexStyle] (A4) at (6, 0) {$A_4$}; 
      \node[VertexStyle] (Y) at (3, 2) {$Y$}; 
      \node[circle, dashed, draw] (Z) at (3, -2) {$\bZ$};
      
     \draw [->, >=stealth', thick=2] (A1) to [out=-50, in=-150] (A3);

      \Edges(A1, Y) \Edges(A1, A2) \Edges(A2, Y) \Edges(A3, Y) \Edges(A4, Y)
      \Edges(Z, A1) \Edges(Z, A2) \Edges(Z, A3) \Edges(Z, A4)
      \Edges(Z, Y)
      
    \end{tikzpicture}
  \end{center}
  \vspace{-.25in}
  \caption{Directed Acyclic Graph in the Presence of Causal Relations
    among Treatments.} \label{fig:causaltreat}
\end{figure}
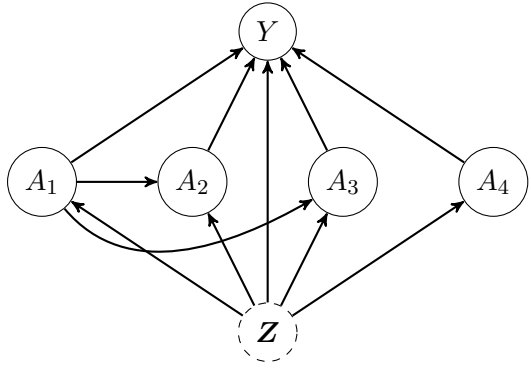

Next, consider a different case shown as the DAG in
Figure~\ref{fig:causaltreat}, in which one treatment causally affects
other treatments.    In this case, we may focus on estimating the causal effects of
$(A_2, A_3, A_4)$ conditional on $A_1$.   We assume the following   model for the outcome
variable,
\begin{eqnarray*}
 \E\{Y_i(\ba) \mid Z_i\} \ = \ \beta_0+ \sum_{j=1}^4 \beta_j A_{ij}+ \sigma Z_i.
\end{eqnarray*}
The joint distribution of $\bA_i$ and $Z_i$ under  Figure~\ref{fig:causaltreat}
is given by
$p(Z_i) p(A_{i1} \mid Z_i)p(A_{i2} \mid A_{i1}, Z)p(A_{i3}\mid A_{i1},
Z_i) p(A_{i4} \mid Z_i)$.  This factorization is identifiable from the
observed data \citep{allman2009identifiability}.  
Then, even when 
$\sigma$ is unknown, we can identify the parameters in the outcome model so long as
$\E(Z_i \mid A_{i1},A_{i2},A_{i3},A_{i4})$ is not linear in
$(A_{i1},A_{i2},A_{i3},A_{i4})$. Using these estimated parameters, we can obtain the estimates for the causal effects.

\subsection{Nonparametric Approach}

In the causal inference literature, many scholars first consider the
problem of nonparametric identification by asking whether or not
causal effects can be identified without making any modeling
assumption.  Only after the nonparametric identification of causal
effects is established, researchers proceed to their estimation and
inference.  \citet{cox:donn:11} regard this approach as a general
principle of applied statistics.  They state,
\begin{quote}
  {\it If an issue can be addressed nonparametrically then it will often be
  better to tackle it parametrically; however, if it cannot be
  resolved nonparametrically then it is usually dangerous to resolve
  it parametrically.} (p. 96)
\end{quote}

\begin{figure}[t]
  \begin{center}
    \tikzstyle{VertexStyle} = [shape = circle, minimum width = 2ex, draw]
    \tikzstyle{EdgeStyle}   = [->,>=stealth']      
    \begin{tikzpicture}[scale=1] 
      \SetGraphUnit{2}
      \node[VertexStyle] (A1) at (0, 0) {$A_1$}; 
      \node[VertexStyle] (A2) at (2, 0) {$A_2$}; 
      \node[VertexStyle] (Am) at (6, 0) {$A_m$}; 
      \node[VertexStyle] (Y) at (3, 2) {$Y$};
      \node[VertexStyle] (W) at (0, -2) {$\bW$};
      \node (dots) at (4, 0) {$\cdots$};
      \node[circle, dashed, draw] (Z) at (4, -2) {$\bZ$};
      \Edges(A1, Y) \Edges(A2, Y) \Edges(Am, Y)
      \Edges(Z, A1) \Edges(Z, A2) \Edges(Z, Am) \Edges(Z, Y)
      \Edges(W, A1) \Edges(W, A2) \Edges(W, Am)
    \end{tikzpicture}
  \end{center}
  \vspace{-.25in}
  \caption{Directed Acyclic Graph for the Instrumental Variable Approach.} \label{fig:DAGiv}
\end{figure}
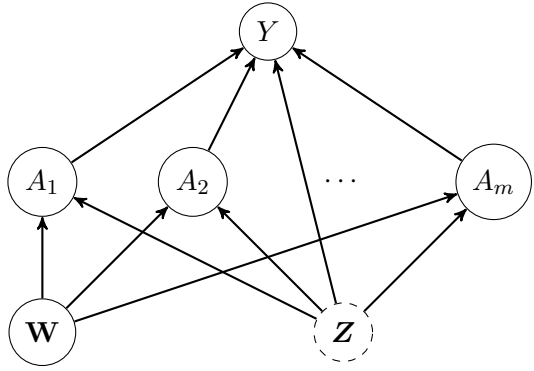

To enable the general nonparametric identification of causal effects
in the current setting, we must introduce auxiliary variables.
\citet{d2019multi} considers the use of proxy variables.  Here, we
examine an approach based on instrumental variables.
Figure~\ref{fig:DAGiv} presents the DAG for this approach where $\bW$
represents a set of instrumental variables.  Instrumental variables
have the property that they are not affected by the unobserved
confounders $\bZ$ and influence the outcome $Y$ only through the
treatments $\bA$.

For the sake of simplicity, we begin by considering the following
separable model for the outcome,
\begin{eqnarray*}
\E\{Y_i(\ba) \mid \bZ_i\} \ = \ q(\ba)+ r(\bZ_i),
\end{eqnarray*}
where $\E\{r(\bZ_i)\}=0$ without loss of generality.  Since the
instrumental variables satisfy
$\E\{r(\bZ_i) \mid \bW_i\}=\E\{r(\bZ_i)\}=0$, we obtain,
\begin{eqnarray}
\label{eqn::iv-nonpara} \E(Y_i \mid \bW_i)&=&  \E\{q(\bA_i) \mid
                                              \bW_i\} \ = \ \sum_{\ba \in
                                              \cA} q(\bA_i = \ba) p(\bA_{i}=\ba \mid \bW_i).
\end{eqnarray}
Since we can identify $\E(Y_i \mid \bW_i)$ and $p(\bA_i\mid \bW_i)$
from the observed data, the causal effects are identifiable if we can
uniquely solve $q(\cdot)$ using equation~\eqref{eqn::iv-nonpara}.
Suppose that all the treatments are binary and the instrumental
variable is discrete with $L$ levels.  Since there are $2^m$
parameters in $q(\ba)$, equation~\eqref{eqn::iv-nonpara} implies that
the identification requires the $2^m\times L$ matrix
$\{ p(\bA_i \mid \bW_i) \}$ to be full-rank.  This condition is
analogous to the overlap assumption discussed earlier and can be
checked using the observed data.  The proposed approach here, however,
requires the instrumental variables to have more than $2^m$
levels. When $m$ is large, it may be difficult to find instrumental
variables that satisfy this condition.

The deconfounder method is closely related to the control function
methods developed in the econometrics literature.  The control
function is a variable that, when adjusted for, renders an otherwise
endogenous treatment variable exogenous \citep[see
e.g.,][]{wooldridge2015control}.  \citet{imbens2009identification}
consider the nonparametric identification of the following
nonseparable triangular system of equations (as before, we omit
observed pre-treatment confounding variables for simplicity),
\begin{eqnarray}
  \label{eqn::ivY} Y_i & = & s_1(A_i, Z_i), \\
   \label{eqn::ivA}  A_i & = & s_2(W_i, U_i)
\end{eqnarray}
where $Z_i$ and $U_i$ are unobserved, $A_i$ is the endogenous treatment
variable of interest, $W_i$ is the instrumental variable with
$W_i \ind (Z_i, U_i)$, and $s_2(\cdot, \cdot)$ is a strictly
monotonic function of $U_i$.  When $A_i$ is a vector and $U_i=Z_i$,
equations~\eqref{eqn::ivY}~and~\eqref{eqn::ivA} become identical to
the setting of the deconfounder method.
\citeauthor{imbens2009identification} show that the control function
$C_i$ is given by the cumulative distribution function of $A_i$ given
$W_i$, i.e., $C_i = F_{A \mid W}(A_i, W_i)$.  Like the substitue
confounder, the control function unconfounds the treatment variable,
i.e., $Y_i(a) \ind A_i \mid C_i$. This is because $C_i$ is a
one-to-one function of $U_i$, and $A_i$ depends only on $W_i$
conditional on $U_i$.

It is important to emphasize that the control function methodology
requires the overlap assumption that the support of the marginal
distribution of the control function, i.e., $p(C_i)$, is the same as
the support of the conditional distribution, i.e., $p(C_i \mid A_i)$.
However, unlike the case of the deconfounder method, the control
function is not a function of the treatment variable, making this
overlap assumption more likely to be satisfied.

In sum, the nonparametric identification of causal effects in the
current settings requires the existence of auxiliary variables.  Here,
we consider an approach based on instrumental variables.  Even when
such instrumental variables are available, certain overlap assumptions
are needed.  This point is also clearly shown for the control function
methods that are closely related to the deconfounder method.  As we
discussed, the overlap assumptions required for these instrumental
variable methods are less stringent than those required for the
deconfounder method.

\subsection{Stochastic Intervention Approach}

Our discussion has identified the overlap assumption as a main
methodological challenge for the deconfounder method.  Because the
estimated substitute confounder itself is a function of treatment
variables, conditioning on the particular treatment values alters the
support of its distribution.  The parametric and nonparametric
approaches introduced above address this problem through the reliance
on modeling assumptions and the use of instrumental variables,
respectively.

The final approach we consider is to change the causal quantities of
interest using the idea of stochastic intervention.  Instead of
comparing two sets of fixed treatment values, we propose to contrast
the two different distributions of treatments.  In the movie
application of the original article, one may be interested in
comparing the revenue of a film featuring a typical cast for action
movies with that featuring common actors for Sci-Fi movies.
Stochastic intervention is a useful approach especially in the
settings where inferring the average outcome under the fixed treatment
values is difficult.  For example, \citet{gene:07} applies it to
mediation analysis, while \citet{hudg:hall:08} propose an experimental
design with stochastic intervention to identify spillover effects.
More recently, \citet{kennedy2019nonparametric} considers the
incremental interventions that shift propensity score values to avoid
overlap assumption.

Specifically, we focus on the average causal effects of distributions 
of treatments rather than the effects of treatments themselves.
\begin{equation}
  \delta(p_1, p_0) \ = \ \E\left\{\int Y_i(\ba) p_1(\bA_i = \ba) d\ba  - \int Y_i(\ba)
  p_0(\bA_i = \ba) d\ba \right\}  
\end{equation}
where $p_1$ and $p_0$ are the pre-specified distributions of
treatments to be compared.  Various distributions can be selected for
comparison.  For example, we may compare the conditional distributions
of treatments given the different values of observed covariates, i.e.,
$p_1(\bA_i \mid \bX_i = \bx_1)$ and $p_0(\bA_i \mid \bX_i = \bx_2)$.
Moreover, if factors are interpretable, then we may choose the
conditional distributions given some specific values of the factors,
i.e., $p_1(\bA_i \mid \bZ_i = \bz_1)$ and
$p_0(\bA_i \mid \bZ_i = \bz_2)$.  Topic models in the analysis of
texts and ideal point models in the analysis of roll calls are good
examples of interpretable factor models
\citep{blei:ng:jord:03,clin:etal:04}.

In the current setting, we may use the following estimator,
\begin{equation}
  \hat\delta(p_1, p_0) \ = \ \sum_{i=1}^n Y_i
  \frac{p_1(\bA_i) - p_0(\bA_i)}{\hat{p}(\bA_i \mid \bZ_i)} \label{eq:estimator}
\end{equation}
where $\hat{p}(\bA_i \mid \bZ_i)$ is the estimated factor model.  For
this estimator, the required overlap assumption is that the support of
$p_j(\bA_i)$ is a subset of the support of $p(\bA_i \mid \bZ_i)$ for
$j=0,1$.  Researchers can choose $p_1(\bA_i)$ and $p_0(\bA_i)$ so that
this overlap assumption is satisfied.  Furthermore, although the
deconfounder method is not applicable when one treatment causally
affects another, under the stochastic intervention approach one could
model causal relationships among treatments by specifying
$p(\bA_i \mid \bZ_i)$ provided that the model is identifiable. An
example of such case is given in Figure~\ref{fig:causaltreat}.

\section{Concluding Remarks}

The article by Wang and Blei is an important contribution to the
causal inference literature because it opens up a new research
frontier.  The authors study a relatively unexplored question of how
to infer the causal effects of many treatments in the presence of
unobserved confounders.  The deconfounder method provides a novel and
yet intuitive approach using familiar statistical models.  A key
insight is that under certain assumptions, the factorization of
treatments can yield a substitute confounder as well as a practically
useful diagnostic tool for checking the validity of the resulting
substitute confounder.

Although the deconfounder method has advantages, as first pointed out
by \citet{d2019multi} and further elaborated in this commentary, the
method is not free of limitations.  In particular, it cannot achieve
nonparametric identification without additional restrictions.  We
emphasized the violation of the overlap assumption due to the fact
that the estimated substitute confounder is a function of observed
treatments.  Wang and Blei consider some restrictions on the outcome
model that may overcome this limitation and enable identification.
However, such restrictions may severely limit the applicability of the
deconfounder method.  More research is needed in order to investigate
the consequences of these restrictions in practical settings.

We discussed three alternative approaches to the methodological
problems of the deconfounder method.  The first approach is based on
parametric assumptions and extend the data analysis conducted in the
original article.  The second approach relies upon the use of
instrumental variables and is related to the control function
literature in econometrics.  The final approach considers an
alternative causal estimand based on stochastic intervention, which is
particularly useful in the settings with high-dimensional treatments.
We expect and hope that many researchers will follow up on the work of
Wang and Blei and develop new methods for estimating the causal
effects of multiple treatments in observational studies.

\clearpage
\spacingset{1.4}
\pdfbookmark[1]{References}{References}
\bibliographystyle{pa}
\bibliography{MultipleCauses-ref,my}

\end{document}